\documentclass[structabstract]{aa}  
\usepackage{url}
\usepackage{graphicx}
\usepackage{txfonts}
\begin{document}
\title{Detection of Extended Molecular Gas in the Disk of the LSB Galaxy Malin~2}


   \author{M.~Das\inst{1,2},
          F.~Boone\inst{3},
          F.~Viallefond\inst{3},
          }

   \institute{Raman Research Institute, Sadashivanagar,
              Bangalore, 560080, India\\
              \email{chandaniket@gmail.com}
         \and
             Birla Institute of Technology and Science - Pilani, Hyderabad Campus, 
Jawahar Nagar, Shameerpet Mandal, Hyderabad, 500078, India
         \and
             Observatoire de Paris, LERMA,
             61 avenue de l'Observatoire, F-75014 Paris, France\\
             \email{frederic.boone@obspm.fr, fviallef@maat.obspm.fr}}

\date{Received ...; accepted ...}

 
  \abstract
   {} 
   {Our goal is to see if there is molecular gas extending throughout 
the optical low surface brightness disk of the galaxy Malin 2.}
   {We used the heterodyne receiver array (HERA) mounted on the IRAM 30m telecope
to make deep observations at the frequency of the CO(2--1) line at nine
different positions of Malin~2. With a total observing time of 11 hours 
at a velocity resolution of 11 km/s we achieve a sensitivity level of ~1 mK.}
   {We detect CO(2-1) line emission from Malin~2.
The line is detected in four of the nine HERA beams; a fifth beam shows a marginal detection. 
These results not only confirm that there is molecular gas in the
disk of Malin 2, but they also show that it is spread throughout the inner 34~kpc radius
as sampled by the observations of the galaxy disk. The mean molecular gas surface density 
in the disk is $1.1\pm0.2~M_{\odot}~pc^{-2}$ and the 
molecular gas mass lies between the limits $4.9\times10^{8}$ to $8.3\times10^{8}~M_{\odot}$.
The observed velocity dispersion of the molecular gas is higher ($\sim 13$\,km\,s$^{-1}$) than 
in star forming galactic disks. This could explain the disk stability and its low star 
formation activity.} 
  {}


   \maketitle
%

\section{Introduction}

Low Surface Brightness (LSB) galaxies are the most unevolved class of galaxies
in our nearby Universe (Impey \& Bothun \cite{ImpeyBothun1997}). They are 
optically dim with diffuse stellar disks (Auld et al. \cite{auld.etal.2006}), 
massive HI gas disks (O'Neil et al. \cite{oneil.etal.2004}; 
Matthews, van Driel, Monnier-Ragaigne \cite{matthews.etal.2001}) but have low star formation rates 
compared to regular spiral galaxies (McGaugh \cite{McGaugh.1994}). They are halo dominated galaxies 
(de Blok \& McGaugh \cite{deblok.etal.1996}; Kuzio de Naray, McGaugh \& de Blok \cite{KuziodeNaray.etal.2008}; 
Coccato et al. \cite{Coccato.etal.2008}); this may account for the weak spiral arms and small bar perturbations
observed in these galaxies (Mihos, de Blok \& McGaugh \cite{Mihos.etal.1997}; 
Mayer \& Wadsley \cite{Mayer.Wadsley.2004}). Although the most commonly observed LSB galaxies are 
the dwarf LSB galaxies (Sabatini et al. \cite{Sabatini.etal.2003}), a significant fraction of 
LSB galaxies are
large spirals having prominent bulges (Beijersbergen, de Blok \& van der Hulst \cite{Beijersbergen.etal.1999}).
These giant LSB (GLSB) galaxies have extended LSB disks that are poor in star formation
and dust (Rahman et al. \cite{Rahman.etal.2007}; Hinz et al. \cite{Hinz.etal.2007}).
The bulge dominated GLSB galaxies often show
AGN activity (Schombert \cite{Schombert.1998}; Das et al. \cite{Das.etal.2009}). 

Even though the optical properties of LSB galaxies have been investigated in great depth, not much
is known about their molecular gas content. This is important as a knowledge of the cold gas distribution in
LSB galaxies will help us understand star formation processes in these galaxies. Surveys of LSB galaxies 
show that they have fairly massive HI disks that may be more than twice the size of the optical disk 
(de Blok et al. \cite{deblok.etal.1996}; Pickering et al. \cite{pickering.etal.1997};
Das et al. \cite{Das.etal.2007}). In this paper we examine the molecular gas distribution in a GLSB galaxy 
and see how it relates to the overall star formation in its' disk. Molecular gas has been detected in only 
a handful of LSB galaxies (O'Neil, Hofner \& Schinnerer \cite{oneil.etal.2000}; 
Matthews \& Gao \cite{matthews.gao.2001}; O'Neil, Schinnerer \& Hofner \cite{oneil.etal.2003};
Matthews et al. \cite{matthews.etal.2005}; Das et al. \cite{Das.etal.2006}). In most cases the galaxies
were large spirals with extended optically dim disks. The low detection rate of
molecular emission from LSB galaxies is probably due to several factors related to the poor star formation
rate in these galaxies (e.g.~de Blok \& van der Hulst \cite{deblok.vanderHulst.1998}); factors such as  
the lower dust content, lower metallicity and the lower surface denity of cold, neutral gas in these galaxies.
All of these properties lead to a slower rate of gas cooling and molecule formation.  For the few galaxies 
where molecular gas has been detected, not much is known about the gas extent and distribution. Such 
information is important if we want to understand star formation and disk evolution in LSB galaxies. 

To investigate the molecular gas and star formation in GLSB galaxies we studied the CO distribution in 
a galaxy where molecular gas has been detected, F568-6 or Malin~2 as it is widely known
(Das et al. \cite{Das.etal.2006}). It is a nearly face-on GLSB galaxy at a distance of 201~Mpc.
It has a prominent bulge and a very extended LSB disk. Its parameters are summarised in Table~1. 
There are several localized star forming regions distributed over its inner disk.
Its metallicity is one third of the solar metallicity in value which is 
relatively high for an LSB galaxy (McGaugh \cite{McGaugh.1994}). 
The CO observations of Malin~2 were conducted using the HERA instrument mounted on the 
30~m IRAM telescope. Our 
main aim was to examine the molecular gas distribution; determine its extent, 
total gas mass and surface density. 
 

\section{Observations}

During March 2007 we observed the CO(2--1) line in Malin~2 with the HERA beam array 
(Schuster et al. \cite{Schuster.etal.2004}) on the IRAM 30m
telescope at a fequency of 220.372~GHz. We specifically used this array as it has a 
wide field of view and good sensitivity. HERA is made of 9 receivers
in a $3\times3$ array spaced by $24^{\prime\prime}$ on the sky. The backend used was the
Wideband Line Multiple Autocorrelator (WILMA). The total bandwidth was 930~Hz; it was made up of
465 channels of 2MHz each. The typical system temperatures $T_{sys}$ were in the range 200-250 K for 8 
receivers; one receiver had a systematically higher $T_{sys}$ in the range 350-450K. The
mean FWHM of each of the nine beams is $11.7^{\prime\prime}$. As our main goal was to detect 
the CO emission line from the
disk of Malin~2, we kept the array fixed on the sky in the standard  pointed mode with the central 
beam pointed to the galactic nucleus; the field was tracked to get deep integrations at fixed
positions in the disk. This enabled us to achieve good sensitivity 
but was at the cost of undersampling the source.  We obtained deep
observations at nine positions of the galaxy with a total on source integration time of
11 hours (Figure~1). The observations were frequency switched. The intrinsic velocity resolution 
was 2.7~km~s$^{-1}$; the data was then smoothed using a Hanning squared function.

We reduced the data using the CLASS software of the GILDAS package \footnote{\url{http://www.iram.fr/IRAMFR/GILDAS}}
by fitting a first order baseline to all spectra within a window going
from -400 to +400\,km\,s$^{-1}$ about the galaxys' systemic velocity of 13830~km~s$^{-1}$. This
window was the same for all nine spectra. The noise 
level is not the same for all receivers; the lowest is 0.7mK
and the highest is 1.6 mK in the 10.9 km/s channels. The noise is also not uniform across the band.
Hence we computed the baseline and the noise in the -400 +400 window using emission free channels only. 
All the details are given in Table~2. The conversion factor from K to Jy is 8.6~Jy~K$^{-1}$. 
To convert from the antenna temperature 
scale $T_A^*$ to the main beam temperature $T_{mb}$ we multiplied $T_A^*$ by the factor
$F_{eff}/B_{eff}~=~1.67$ where $F_{eff}$ is the forward beam efficiency and $B_{eff}$ is the
main beam efficiency.

   \begin{table}
      \caption[]{Galaxy Parameters - Malin~2}
         \label{table1}
     $$ 
         \begin{array}{p{0.35\linewidth}ll}
            \hline
            \noalign{\smallskip}
            Parameter      &  Value & Reference \\
            \noalign{\smallskip}
            \hline
            \noalign{\smallskip} 
            Other Names  & F568-6  & NED \\
            Distance & 201~Mpc & NED \\
            Heliocentric Velocity & 13830~km~s^{-1} & NED \\
            Position (epoch 2000) & 10^{h}39^{m}52^{s}.5, +20^{\circ}50^{\prime}49^{\prime\prime} & 2MASS \\
            Size ($D_{25}^{\prime}$) & 1.67 & (a) \\
            Position angle & 70^{\circ} & NED \\ 
            Inclination   & 38^{\circ} & (a) \\
            HI Linewidth ($W_{20,corr}$) & 637~km~s^{-1} & (b) (c) \\
            HI Mass & 4.2\times10^{10}~M_{\odot} & (b) \\
            Dynamical Mass & 2.5\times10^{12}~M_{\odot} & (b) \\
             (R~$<107$~Kpc) &                           &     \\
            Disk Central Brightness & 22.1~mag~arcsec^{-2} & (b) (d) \\
            Disk Scale Length & 18.8~Kpc & (b) \\
            Total R Band Magnitude & -23.6~mag & (b) \\
            \noalign{\smallskip}
            \hline
         \end{array}
     $$ 
\begin{list}{}{}
\item[$^{\mathrm{a}}$] Matthews et al. (2001)
\item[$^{\mathrm{b}}$] Pickering et al. (1997)
\item[$^{\mathrm{c}}$] Corrected for inclination
\item[$^{\mathrm{d}}$] Not corrected for extinction

\end{list}
   \end{table}

\section{Results}

We have detected CO(2-1) emission from several positions in the disk as well
as from the center of Malin~2. In 
the following paragraphs we present our results and discuss their implications.\\

\noindent
{\bf 1.~CO(2-1) Detections~:~}Figure~2 shows the CO(2-1) emission spectra observed from 
nine locations across Malin~2. The offsets from the galaxy center 
are indicated in each box. CO(2-1) emission was detected from four out of nine positions
at line intensities above $3\sigma$. The line at (-24, 24) is a a hint of emission rather
than a sure detection. At (0, 0) the line is broad but is a $3\sigma$ detection.
We estimated the line parameters (flux, width and central velocity)  
by fitting a gaussian to each spectrum. We determined the noise level for each of the nine 
positions. The results are listed in Table~2 and the best fit gaussians are overlaid on the 
spectra in Figure~2. 

\noindent
{\bf 2.~Comparison with previous CO observations~:~}
Molecular gas has been detected earlier in this galaxy by Das et al.
(2006) with the IRAM 30m telescope and the BIMA array. However, due to a
correlator setup problem the BIMA map was incomplete, it was covering a
velocity range corresponding to CO(1--0) line emission from the east side
of the galaxy only. In addition, the 30m telescope was pointed toward two
directions only at 7$''$ and $\sim$35$''$ east of the nucleus (see
Fig.\,1). Although the source was observed simultaneously at 115\,GHz and
230\,GHz with the 30m telescope, only CO(1--0) emission was detected; 
the CO(2--1) line was not detected from either positions. 
Our present  observations cover the CO emission over a wider
velocity and spatial extent than these previous observations and
hence give a better idea of the distribution of molecular gas in the inner
disk of Malin 2. These
observations are also the first detections of the CO(2--1) line in Malin 2.
Our detection lies below the detection limit of the older observations of 
Das et al. (2006). Their CO(2--1) emission spectrum had a noise (rms) 
of 2.6~mK whereas our line detections have a peak of 2  to 3~mK (Figure~2).

Near the galaxy center, where there is overlap with the previous CO
detection,  the CO line 
velocities are in the same direction and are similar in shape and in width.
Close to the center of the galaxy the CO(1--0) line as detected by 
Das et al. (2006) is broad ($\sim$200\,km\,s$^{-1}$) and asymmetric with the 
red side being slightly more prominent than the blue side. 
The CO(2­-1) line detected at the center of the galaxy by our new HERA
observations is $243\pm76~km~s^{-1}$ wide and at a centroid position
offset by $77\pm34~km~s^{-1}$ relative to the systemic velocity given in Table~1.
Investigating the cause of this offset would require interferometric observations.
As the beams and pointing directions are different (23$''$ for the CO(1--0) line and
11$''$ for the CO(2--1) line) the fluxes cannot be directly compared. But if the molecular 
gas were uniformly distributed we would expect a line ratio in temperature scale close in the
range of values found by Braine \& Combes (1992) with the same telescope
i.e., 0.4 to 1.2. We checked this for the HERA central beam whose
pointing direction is the closest to the central pointing of Das et al (2006). A rough comparison 
of the brightness in $mK~km~s^{-1}$ gives
$I_{\rm CO(2-1)}/I_{\rm CO(1-0)}$~=~0.43/1.01~=~0.43 
(where the CO(2--1) line intensity from the HERA central pointing
is 0.43 and the CO(1--0) intensity from IRAM is 1.01). This value is at the lower end of the line ratio range
for most galaxies but given the uncertainties (e.g.~gas distribution), the value shows that the CO(2--1) 
line intensity we measure is consistent with the previous CO(1--0) detection.

\noindent
{\bf 3.~Comparison with HI observations~:~}
High sensitivity HI VLA observations of Malin 2 were presented by Pickering et al. (1997). These reveal the
distribution of the HI gas and its kinematics at a spatial resolution of $\sim$19$''$. The HI disk 
is lop-sided, the HI emission being significantly more prominent and extending farther out on 
the western side of the nucleus than on the eastern side. The central region is essentialy devoid 
of HI gas. Although we have only a partial view of the molecular disk due to our under-sampled data 
its properties have similitude with  HI. 
Based on the HI velocity field, the centroid velocity of the HI line profiles at the 
positions (24,0), (-24,0) and (-24,24), respectively +90, -150 and -110 km\,s$^{-1}$ relative to the 
systemic velocity, are in excellent agreement to those for the molecular gas (Tab. 2, col {\it e}). 
Hence the HI and molecular gas appear to share a common velocity assymmetry when considering the 
positions (24,0) and (-24,0). 

The higher spatial and spectral resolutions in CO allow to better constrain 
the velocity dispersion of the gaseous component. The CO profile with the
narrowest line width is at (-24,0). Given the spectral response, this leads to 
a deconvolved velocity dispersion of 13 km\,s$^{-1}$. It is the best constraint 
that we can get for the velocity dispersion of the gas $\sigma_{gas}$ in the disk of
Malin 2 given the facts that {\it a)} it is obtained in a direction close to 
the major axis where beam smearing effects due to the rotation of the disk are minimum 
and {\it b)} we do not observe the blue-shifted wing or 
secondary velocity component that Pickering et al. discovered and had to blank 
to determine and analyze the rotation curve.  These authors interpreted this
secondary component as high velocity clouds, gas infalling into 
the disk southwest of the nucleus. It is best seen $\sim$20'' southwest of the nucleus 
where it is blue-shifted by as much as $\sim$100\,km\,s$^{-1}$. Given the correspondance
with HII regions in that direction and the fact that there is a relatively moderate HI column
density there they suggested that this infall induces star formation 
(Wyder et al. \cite{Wyder.etal.2009}). 
We do not detect CO emission at (-24,-24) which is only $\sim$6'' further away in the southwest.
It could be that the disk component in that region is depeleted not only in HI but also 
in molecular gas. CO observations spatially fully sampled would be required to confirm 
this hypothesis.

In Table 2, the maximum CO(2--1) line velocities are reached in the offsets (24,0) and (-24,0) which is 
consistent with the
position angle being close to horizontal $\sim70^{\circ}$ (2MASS images; Jarrett et al. \cite{Jarrett.etal.2003}).
We note a velocity asymmetry between the emission at (24,0) and that at (-24,0), which could be due to infalling 
gas or a lopsided potential. We can estimate the rotational velocity from these lines.
Assuming an inclination of 38$^{\circ}$ (Matthews et al. \cite{matthews.etal.2001})
it is of the order ($\sim 150/\sin(i)\,km\,s^{-1}$ which implies a dynamical mass within 
the 24$''$ (i.e. 23\,kpc) radius of $\sim3\times10^{11}~M_{\odot}$.

The broad line at the center has a width close to twice the velocities at  
(24,0) and (-24,0) offsets. This could be due to the rotation curve turning point 
being inside the central beam, i.e., inside a radius of 6$''$ (6\,kpc).
This central region which is essentially devoid of HI is where there is the strongest 
emission in CO; higher angular resolution CO observations are required to determine the 
rotation curve in that region. Would the curve continue to steadily decrease up to 
the nucleus that would point to a model where the gas in the central bulge region 
is kinematically hot, e.g. because of gas infall to feed the AGN.

\noindent
{\bf 4.~Molecular Gas Mass:~}
We derived the molecular gas surface densities for each line detection and the results are summarised in Table~2. 
The column densitites are derived from the line fluxes using 
$N(H_{2})~=~X~(1/0.9)~\int~T_{mb}[CO(2-1)]~dv$; where I(2-1)/I(1-0)~=~0.9
(Braine \& Combes \cite{Braine.Combes.1992}) and the $CO$ to $H_{2}$ 
conversion factor is given by $X=2\times 10^{20}~cm^{-2}(K~km~s^{-1})$
for cool virialized molecular clouds (Dickman et al. \cite{Dickman.etal.1986}). The
value of $X$ is somewhat uncertain for LSB disks, given their low metallicities and our poor 
knowledge of the ISM conditions in these galaxies. However, for a first estimate we    
used this value to determine the surface densities of molecular gas $\Sigma_{mol}$ 
in Malin~2 (Col. f Table~2). Assuming the eight directions observed in the disk form a 
distribution of surface brightness values which is representative for the disk of 
Malin~2 up to a radius of $40^{\prime\prime }$, the mean CO(2--1) brightness  
is $\sim0.15~K~km~s^{-1}$.  
The central region has a beam averaged surface density that is 3 times higher than
the mean brightness of the disk. The lower limit for $M_{mol}$ is $\sim~4.9\times10^{8}M_{\odot}~$ and is 
given by the sum of the four detections in column~{\it f} of Table~2 multiplied by the beam solid angle.
To determine the upper limit for $M_{mol}$, we used the noise   
(see Table~2, column~{\it a}) to derive upper limits for the molecular gas surface densities for the regions 
where there were no detections including (-24,24) where there is a hint of emission. The upper limit to the 
total mass of molecular gas within $R~<~40^{\prime\prime}$ 
is thus $M_{mol}\sim~8.3\times10^{8}~M_{\odot}$.
If the molecular disk does not extend beyond $40^{\prime\prime}$, then the molecular gas mass 
corresponds to $\sim1.2 - 2$\% of the HI gas mass in the galaxy. This is similar to that observed in normal spiral 
galaxies that have molecular gas masses typically a few percent of the total HI mass.

\noindent
{\bf 5.~Star Formation threshold~:~}
The molecular gas disk in Malin~2 is extended and relatively massive with a significant amount of
molecular gas but its star formation activity is low compared to normal spirals 
(Wyder et al. \cite{Wyder.etal.2009}). For galactic disks star formation appears to be controled by the
onset of gravitational instabilities (Kennicutt \cite{Kennicutt.1989}). A simple single-fluid 
Toomre disk stability model predicts threshold densities in agreement with the observations. 
This critical density is given by
$\Sigma_{crit}~=~\alpha\kappa\sigma_{gas}/3.36~G$
where $\kappa$ is the epicyclic frequency and $\sigma_{gas}$ the velocity dispersion of the gas.
From a sample of spiral galaxies Kennicutt found $\alpha$ $\simeq$ 0.69. 
In his study, to evaluate  $\alpha$, 
Kennicutt assumed that all the galaxies in his sample have approximately the same velocity dispersion,
$\sim6 km~s^{-1}$, and that it is independent of the radial distance within a galaxy. This is a good 
approximation given what is known for the galaxies of his sample. Hence the prediction is that, 
to have disk galaxies such as the LSBs immune to star formation, the gas surface density should 
be below this critical density.
 
Pickering et al. (\cite{pickering.etal.1997}) determined rotation curves for 4 GLSBs. These allow 
them to determine the epicyclic 
frequency necessary to obtain the radial profiles of $\Sigma_{crit}$. They show that, by assuming that
$\alpha = 0.69$ is valid for these LSB galaxies and that $\sigma_{gas}=10 km~s^{-1}$, the critical 
density is, in some regions, close or lower than the observed HI surface density, in spite of the fact 
that there is no conspicuious star formation activity there. This leads them to postulate
that $\sigma_{gas}$ must be larger than the assumed value. 

In the specific case of Malin 2 Pickering et al. find that the 
HI surface density never exceeds the critical density except at the location where HI is 
the brightest (this direction is located a half way between the positions (-24,24) and (-24,0)). 
When considering the 9 directions listed in table 2 only (0,0) coincides with an H$\alpha$ 
source. Using the present CO observations we can examine two questions: 
{\it 1)} does the gas surface density remain below the critical density 
when the contribution from the molecular gas component is taken into account 
as did Kennicutt in his analysis? {\it 2)} is $\sigma_{gas} \le 10 km~s^{-1}$ used to
determine  $\Sigma_{crit}$ adapted in the case of giant LSBs such as Malin 2? 
Column {\it g} in Table 2 gives the HI surface brightnesses. Comparing
these with the molecular surface densities (col. {\it f}), the molecular gas surface 
densities appear to amount to typically 1/2 to 1/3 the HI surface densities,
with the exception for the position (-24,0) where HI largely dominates. Column {\it h} gives 
the total gas surface density and in col. {\it i} this quantity relative to the critical
density. By using $\alpha = 0.69$, for internal consistency, the values in col. {\it h} and {\it i}
are determined using the same hypothesis as those adopted by Kennicutt for the conversion 
factor X and the term Z which accounts for the elements heavier than H. 
For  $\sigma_{gas}$ we adopted the value of $13~km~s^{-1}$ 
that we measured with good S/N at (-24,0) (sect 3.3). Better constrained than the $10~km~s^{-1}$ 
adopted by Pickering et al., this value is still conservative given the fact that 
it is the lowest velocity dispersion (col. {\it d}) amongst the 5 directions where CO 
is detected. Formally, all the values in col.  {\it i} being smaller than unity it could be
concluded that the disk of Malin 2 is sufficiently stable to prevent star formation. However,
the directions (-24,0) and (-24,24) have surface densities close to the threshold preventing 
us to consider this as firm solid results. Also
the observed kinematics in western and north-western parts of Malin 2 appears complex because
of high velocity clouds (HVC). It could be that the star formation activity seen in Malin 2 
is more related to the interaction of these HVCs with the disk than to growth of density
fluctuations originating from instabilities in the self-graviting rotating disk. A fully sampled 
image of the Malin 2 and high signal-to-noise to disentangle different velocity components
and measure the line widthes are necessary to better understand the properties of Malin 2.

   \begin{figure}
   \centering
   \includegraphics[width=8cm,angle=-90]{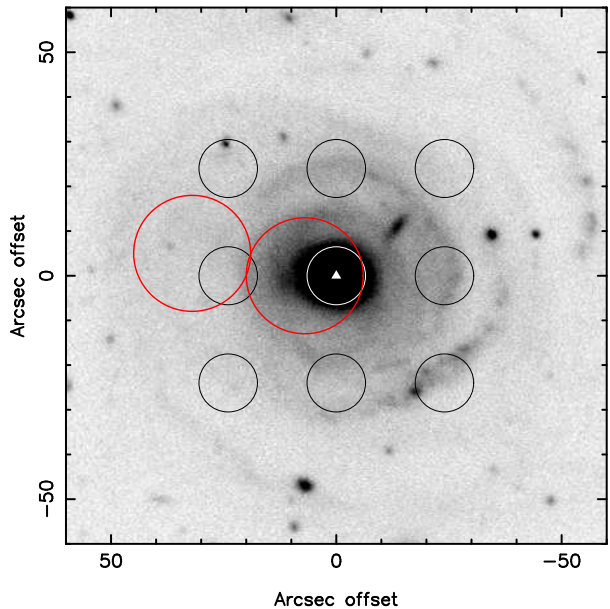}
      \caption{The figure shows the R band image of the galaxy Malin~2 with 
the HERA footprint overlaid as nine black circles. The red circles correspond to
the previous IRAM observations at 115GHz (Das et al. 2006). Each circle has a
diameter equal to the FWHM of the corresponding beam.}
         \label{Fig1}
   \end{figure}
%
   \begin{figure*}
   \centering
   \includegraphics[width=8cm,angle=-90]{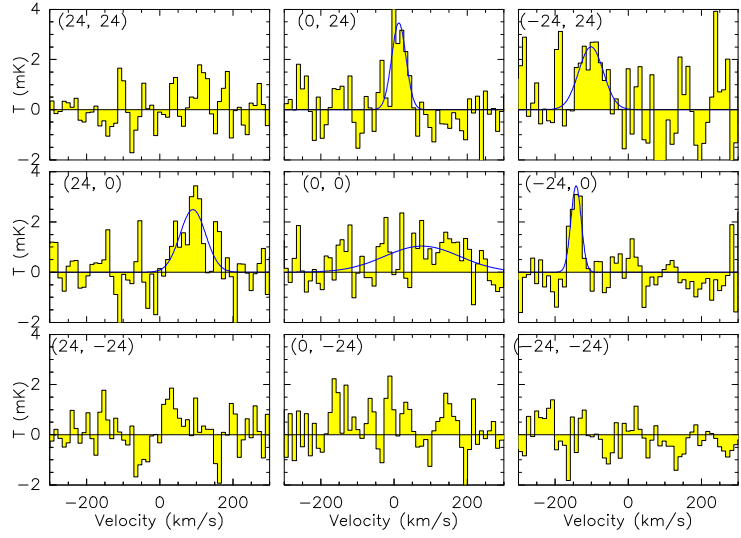}
   \caption{The panel shows the CO(2-1) spectra obtained with the nine HERA receivers. 
The offsets in arcsecond from the galaxy center are marked on the 
top left hand corner of each figure. CO(2-1) emission was detected from five 
locations. The gaussian fits for these detections are overlaid on the spectra. 
The x axes are the velocity offsets from the galaxy systemic velocity ($v_{\rm hel}=13,830~km~s^{-1}$)
and the y axes are the measured antenna temperatures ($T_A^*$) in $mK$.}
    \label{Fig1}%
    \end{figure*}
%
   \begin{table*}
      \caption[]{CO(2-1) fluxes, kinematical parameters, molecular, HI and total gas surface densities, }
\vspace{-3mm}
         \label{table1}
     $$
         \begin{array}{ccccrrrrcccc}
            \hline
            \noalign{\smallskip}
\multicolumn{2}{c}{Offset}    
&  rms~noise  
& {\Delta~I_{CO}}
& \multicolumn{2}{c}{\Delta~V_{FWHM}}   
& \multicolumn{2}{c}{<V>}       
& \Sigma_{\rm mol} 
& \Sigma_{\rm HI} 
&  \multicolumn{1}{c}{\Sigma_{\rm gas}}
&{\frac{\Sigma_{\rm gas}}{\Sigma_{\rm crit}}}\\
\multicolumn{2}{c}{(arcsec)}      
& (mK)     
& (K~km~s^{-1})        
& \multicolumn{2}{c} {(km~s^{-1})} 
& \multicolumn{2}{c}{(km~s^{-1})} 
& (M_{\odot}pc^{-2})  
& (M_{\odot}pc^{-2})  
& (M_{\odot}pc^{-2}) 
& \\
\multicolumn{2}{c}{a}
&b
&c
&\multicolumn{2}{c}{d}
&\multicolumn{2}{c}{e}
&f
&g
&h
&i\\
            \noalign{\smallskip}
            \hline
            \noalign{\smallskip}
+24 & +24  &  0.7    & <~0.12     &\multicolumn{2}{c}{0.39} & \multicolumn{2}{c}{....} & <~0.4 & 2.1 & <~3.0 & ... \\
+24 & +00  &  0.9    & 0.37\pm0.07 &  82~\pm&20  & 91~\pm&~~7  & 1.2\pm0.2                     & 2.5 &     4.8 & 0.7 \\
+24 & -24  &  0.8    & <~0.13     & \multicolumn{2}{c}{0.45}  & \multicolumn{2}{c}{....} & <~0.5 & 1.8 & <~2.8 & ... \\
+00 & +24  &  0.9    & 0.28\pm0.05 &  46~\pm&10  & 14~\pm&~~4  & 0.9\pm0.2                     & 2.7 &     4.6 & 0.7 \\
+00 & +00  &  0.8    & 0.45\pm0.11 & 243~\pm&76 & 77~\pm&34    & 1.5\pm0.4                     & 0.9 &     3.4 &     \\
+00 & -24  &  0.9    & <~0.15      & \multicolumn{2}{c}{0.50} & \multicolumn{2}{c}{....} & <~0.5 & 2.2 & <~3.3 & ... \\
-24 & +24  &  1.6    & \sim0.35\pm0.13 &  79~\pm&41 & -102~\pm&14  & \sim1.2\pm0.4               & 2.2 &  \sim4.4 & \sim0.9 \\
-24 & +00  &  0.9    & 0.20\pm0.04 &  33~\pm&~7  & -142~\pm&~~4& 0.7\pm0.1                     & 4.0 &     5.7 & 0.9 \\
-24 & -24  &  0.7    & <~0.12      & \multicolumn{2}{c}{0.39} & \multicolumn{2}{c}{....} & <~0.4 & 1.8 & <~2.7 & ... \\
            \noalign{\smallskip}
            \hline
         \end{array}
     $$
\begin{list}{}{}
\item[$^{\mathrm{a}}$] $\Delta\alpha$ and $\Delta\delta$ offsets relative to the position of the nucleus $\alpha,\delta$ (J2000): $10^{h}39^{m}52^{s}.5, +20^{\circ}50^{\prime}49^{\prime\prime}$
\item[$^{\mathrm{b}}$] at 10.9 $km~s^{-1}$ velocity resolution.
\item[$^{\mathrm{c}}$] $\Delta I_{\rm{CO(1-0)}}=\int T_{\rm MB} dv$ with $T_{\rm MB}=1.67 T_A^*$. For the non-detections, 
we take $T_A^*$=noise and assume $\Delta V=100~km~s^{-1}$.
\item[$^{\mathrm{d}}$] fitted line widths, not deconvolved by the spectral response. 
\item[$^{\mathrm{e}}$] fitted, relative to the systemic velocity 13830$~km~s^{-1}$.
\item[$^{\mathrm{f}}$] $\Sigma_{\rm mol}~=~2m_{p}N(H_{2})$. Values for X=2.0 10$^{20}~cm^{-2}~K~km~s^{-1}$ (ref. sect. 3.4).
\item[$^{\mathrm{g}}$] from Fig. 3 of Pickering et al. (\cite{pickering.etal.1997}). 
\item[$^{\mathrm{h}}$] $\Sigma_{\rm gas} = (1+Z)~(\Sigma_{\rm HI}+\frac{X}{2.0 10^{20}}\Sigma_{\rm mol})~cos(i).$ 
                       with Z=0.45, X=2.8 10$^{20}~cm^{-2}~K~km~s^{-1}$ and i=38$^{\circ}$.
                       Values in brackets assume $\Sigma_{\rm mol} \ll \Sigma_{\rm HI}$
\item[$^{\mathrm{i}}$] using $\Sigma_{\rm crit}$ from Fig. 10 of Pickering et al. (\cite{pickering.etal.1997}), 
                       re-scaled for $\sigma_{gas}=13~km~s^{-1}$ 
\end{list}
   \end{table*}


\section{Conclusions}

   \begin{enumerate}
      \item We report the first detection of CO(2-1) line emission from the disk of a giant 
LSB galaxy. Our observations reveal the presence of extended molecular gas in the disk of the LSB galaxy
Malin~2. The radial extent of the molecular disk is at least 34~kpc and its mean line brightness
$\sim 0.15 K~km~s^{-1}$ (on the $T_{mb}$ temperature scale). With an angular
resolution of $11.7^{\prime\prime}$ this is $\sim 2/3$ the beam-averaged integrated line intensity
in the central region.
      \item When we compared the molecular and HI gas masses of Malin~2, we found that the
molecular gas fraction is $\sim$1.2~-~2\% of the  HI gas mass. 
At radii $\sim~$15 kpc in the directions where CO is detected, the molecular gas 
fraction is typically 30 to 50\% of the HI gas surface densities.
       \item We estimated the total surface density of the neutral gas and found that it is always
below the critical threshold density for gravitational instability. 
Hence the disk of Malin 2 is overall stable against large scale star formation. 
   \end{enumerate}

\begin{acknowledgements}
We are grateful to the IRAM staff at Pico Veleta for excellent support at the 
telescope. IRAM is supported by INSU/CNRS (France), MPG (Germany) and IGN (Spain).
The authors would also like to thank Alice Quillen for providing the R-band images of Malin~2.
This research has made use of the NASA/IPAC Extragalactic Database (NED) which is operated 
by the Jet Propulsion Laboratory, California Institute of Technology, under contract with the 
National Aeronautics and Space Administration.
\end{acknowledgements}


\begin{thebibliography}{}

  \bibitem[2006]{auld.etal.2006} Auld, R.; de Blok, W. J. G.; Bell, E.; Davies, J. I. 
2006, MNRAS, 366, 1475

  \bibitem[1999]{Beijersbergen.etal.1999} Beijersbergen, M.; de Blok, W. J. G.; van der Hulst, J. M. 
1999, A\&A, 351, 903

  \bibitem[2000]{Braine.Combes.1992} Braine, J.; Combes, F. 1992, A\&A, 264, 433 

  \bibitem[2008]{Coccato.etal.2008} Coccato, L.; Swaters, R. A.; Rubin, V. C.; D'Odorico, S.; 
McGaugh, S. S.  2008, A\&A, 490, 589

  \bibitem[2006]{Das.etal.2006} Das, M., O'Neil, K., Vogel, S.~N., McGaugh, S. 2006, ApJ, 651, 853

  \bibitem[2007]{Das.etal.2007} Das, M.; Kantharia, N.; Ramya, S.; Prabhu, T. P.; McGaugh, S. S.; 
Vogel, S. N. 2007, MNRAS 379, 11

  \bibitem[2009]{Das.etal.2009} Das, M., Reynolds, C. S.; Vogel, S. N.; McGaugh, S. S.; Kantharia, N. G
2009, ApJ, 693, 1300 

  \bibitem[1986]{Dickman.etal.1986} Dickman, R. L.; Snell, Ronald L.; Schloerb, F. Peter 1986, ApJ, 309, 326

  \bibitem[1996]{deblok.etal.1996} de Blok, W.J.G.; McGaugh, S.S.; van der Hulst, J.M. 1996, MNRAS, 283, 18 

  \bibitem[1998]{deblok.vanderHulst.1998} de Blok, W.J.G.; van der Hulst, J.M. 1998 A\&A, 336, 49 

  \bibitem[2007]{Hinz.etal.2007} Hinz, J. L.; Rieke, M. J.; Rieke, G. H.; Willmer, C. N. A.; Misselt, K.; Engelbracht, C. W.; Blaylock, M.; Pickering, T. E.  2007, ApJ, 663, 895

  \bibitem[1997]{ImpeyBothun1997} Impey, C. \&Bothun, G. 1997, ARA\&A, 35, 267

  \bibitem[2003]{Jarrett.etal.2003} Jarrett, T. H.; Chester, T.; Cutri, R.; Schneider, S. E.; Huchra, J. P.  2003, AJ, 125, 525

  \bibitem[1989]{Kennicutt.1989} Kennicutt, R.C. 1989, ApJ, 344, 685

  \bibitem[2008]{KuziodeNaray.etal.2008} Kuzio de Naray, Rachel; McGaugh, Stacy S.; de Blok, W. J. G.
2008, ApJ, 676, 920

  \bibitem[2001]{matthews.etal.2001} Matthews, L. D.; van Driel, W.; Monnier-Ragaigne, D. 
2001, A\&A, 365, 1

  \bibitem[2001]{matthews.gao.2001} Matthews, L.D. \& Gao, Y. 2001, ApJ 549, L191

  \bibitem[2005]{matthews.etal.2005} Matthews, L.D.; Gao, Y.; Uson, J.M.; Combes, F. 2005, AJ, 129, 1849

  \bibitem[2004]{Mayer.Wadsley.2004} Mayer, L. \&  Wadsley, J.  2004 MNRAS, 347, 277

  \bibitem[1994]{McGaugh.1994} McGaugh, S.S. 1994 ApJ, 426, 135

  \bibitem[1995]{McGaugh.etal.1995} McGaugh, S.S.; Bothun, G.D.; Schombert, J.M. 1995 AJ, 110, 573

  \bibitem[1997]{Mihos.etal.1997} Mihos, J. C.; McGaugh, S. S.; de Blok, W. J. G 1997, ApJ, 477L, 79
  
  \bibitem[2004]{oneil.etal.2004} O'Neil, K.; Bothun, G.; van Driel, W.; Monnier Ragaigne, D  2004,
A\&A, 428, 823

  \bibitem[2000]{oneil.etal.2000} O'Neil, K.; Hofner, P. \& Schinnerer, E. 2000, ApJ, 545, L102

  \bibitem[2003]{oneil.etal.2003} O'Neil, K.; Schinnerer, E.; \& Hofner, P. 2003, ApJ, 588, 230

  \bibitem[1997]{pickering.etal.1997} Pickering, T. E.; Impey, C. D.; van Gorkom, J. H.; Bothun, G. D. 1997, 
AJ, 114, 1858

\bibitem[2007]{Rahman.etal.2007} Rahman, Nurur; Howell, Justin H.; Helou, George; Mazzarella, Joseph M.; Buckalew, Brent  
2007, ApJ, 663, 908

  \bibitem[2004]{Schuster.etal.2004} Schuster, K.-F.; Boucher, C.; Brunswig, W.; Carter, M.; Chenu, J.-Y.; Foullieux, B.; Greve, A.; John, D.; Lazareff, B.; Navarro, S.; and 5 coauthors 2004, A\&A, 423, 1171

  \bibitem[2003]{Sabatini.etal.2003} Sabatini, S.; Davies, J.; Scaramella, R.; Smith, R.; Baes, M.; 
Linder, S. M.; Roberts, S.; Testa, V. 2003, MNRAS, 341, 981

  \bibitem[1998]{Schombert.1998} Schombert, J.M. 1998; AJ, 116, 1650
  
  \bibitem[2009]{Wyder.etal.2009} Wyder, T.K.; Martin, D.C.; Barlow T.A.; Foster, K.; Friedman, P.G.; Morrissey, P.;
Neff, S.G. 2009, ApJ, 696, 1834

\end{thebibliography}
\end{document}